\documentclass{jfm}

\usepackage{graphicx}
\usepackage{newtxtext}
\usepackage{newtxmath}
\usepackage{natbib}
\usepackage{hyperref}
\hypersetup{
    colorlinks = true,
    urlcolor   = blue,
    citecolor  = black,
}

\usepackage{caption}
\newcommand{\RomanNumeralCaps}[1]
\linenumbers

\usepackage{xcolor}

\title{Joint Reduced Model for the Laminar and Chaotic Attractors in Plane Couette flow}

\author{Bálint Kaszás\aff{1\corresp{\email{bkaszas@ethz.ch}}}
  \and George Haller\aff{1} }

\affiliation{\aff{1}Institute for Mechanical Systems, ETH Z\"{u}rich Leonhardstrasse 21, 8092 Z\"{u}rich, Switzerland}

\begin{document}
\maketitle

\begin{abstract}
We use the theory of spectral submanifolds (SSMs) to develop a low-dimensional reduced-order model for plane Couette flow in the permanently chaotic regime studied by \cite{kreilos_eckhardt_2012}. Our three-dimensional model is obtained by restricting the dynamics to the slowest mixed-mode SSM of the edge state. We show that this results in a nonlinear model that accurately reconstructs individual trajectories, representing the entire chaotic attractor and the laminar dynamics simultaneously. In addition, we derive a two-dimensional Poincaré map that enables the rapid computation of the periodic orbits embedded in the chaotic attractor.
\end{abstract}



\section{Introduction}
\label{sec:intro}
Certain shear flows, such as pipe (\cite{avila_review_2023}) and plane Couette flow (\cite{eckhardt_TransitionTurbulenceShear2018}) exhibit subcritical transition to turbulence, that results in a coexistence of an extended turbulent state with a stable laminar state. The boundary between the turbulent and laminar behaviors is often called the edge of chaos (\cite{skufca_EdgeChaosParallel2006}). For Couette flow in small periodic domains, this edge of chaos is the stable manifold of an unstable exact coherent state (ECS) (\cite{waleffe_ExactCoherentStructures2001}), called the edge state (\cite{wangLowerBranchCoherent2007}). 

It is known that ECSs, such as fixed points (\cite{gibsonEquilibriumTravellingwaveSolutions2009}), traveling waves (\cite{WEDIN_KERSWELL_2004}), and periodic orbits (\cite{budanur_RelativePeriodicOrbits2017}) play an important role in the turbulent dynamics. Indeed, complex behavior tends to develop through bifurcations of simpler ECSs. In particular, a period doubling cascade (\cite{Moore1983}), or the bifurcation of invariant tori (\cite{Ruelle1971, gollub_1975}) typically precedes the appearance of turbulent flow. 

For these reasons, great effort has been directed towards identifying unstable ECSs in direct numerical simulations (DNS) (\cite{kawaharaPeriodicMotionEmbedded2001, graham_ExactCoherentStates2021, page_norgaard_2024}) and in experiments (\cite{suri_HeteroclinicHomoclinicConnections2019a}). \cite{page_norgaard_2024} found that certain turbulent statistics can be approximated using a combination of a finite set of periodic orbits that approximate the chaotic attractor. However, the weights corresponding to these orbits are not determined a priori. Similarly, \cite{yalnizCoarseGrainingState2021} found a probabilistic Markov chain model based on periodic orbits. Despite these results, periodic orbit theory (\cite{ChaosBook}) has not yet been successfully applied to turbulent flows. 

Developing predictive dynamical models is even more challenging. Among data-driven approaches, linear methods such as dynamic mode decomposition (DMD) (\cite{schmid_DynamicModeDecomposition2010}) are necessarily inapplicable due to the nonlinearizable nature of turbulent flows, as shown by \cite{page_KoopmanModeExpansions2019}. Recently, advances in machine learning have yielded a promising alternative. Specifically, the deep learning-based \texttt{DManD} method (\cite{linot_DeepLearningDiscover2020}) captures turbulent statistics and can also represent ECSs not enforced in their training (\cite{Linot_Graham_2023}). However, deep learning methods generally suffer from a need for large amounts of training data, a time-consuming training process, and an a priori unclear choice of hyperparameters. 

In contrast, reducing a shear flow to a low-dimensional, attracting, and structurally stable invariant manifold in its phase space offers a mathematically exact and robust construction of a reduced-order model. The recently introduced theory of spectral submanifolds (SSMs) (\cite{haller_NonlinearNormalModes2016a, haller2025}) targets low-dimensional attracting invariant manifolds emanating from stationary states, such as ECSs. Specifically, primary SSMs are defined as the smoothest invariant manifolds tangent to a selected spectral subspace of the linearized dynamics at the stationary state. Building on prior work by \cite{cabre_ParameterizationMethodInvariant2003}, 
 \cite{haller_NonlinearNormalModes2016a} established the existence and uniqueness of primary SSMs for attracting fixed points under mild nonresonance conditions. Secondary SSMs of lower smoothness class, mixed-mode SSMs for saddle-type fixed points, and SSMs under general (aperiodic and even random) forcing have been shown to exist. Furthermore, data-driven (\cite{cenedese_DatadrivenModelingPrediction2022}) and equation-driven (\cite{jain_HowComputeInvariant2022}) SSM identification methods have been successfully applied to mechanical vibrations (\cite{axas_FastDatadrivenModel2023}), fluid dynamics (\cite{xu_2024}), and robotics (\cite{Alora2025}), summarized by \cite{haller2025}. 

Apart from providing verifiable mathematical conditions for its applicability, SSM-based model reduction has the advantage over usual manifold learning approaches of providing the dimension of the underlying SSM a priori. This is in contrast to, e.g., autoencoder-based methods that require careful optimization of the latent dimension as a hyperparameter. 


These features enable SSM-based model reduction to capture even chaotic attractors from data, as demonstrated by \cite{liu_2023}. However, no data-driven reduced-order models have been constructed for fluid flows with coexisting nontrivial attractors. 

In this paper, we fill this gap in reduced flow modeling by constructing the slowest three-dimensional (3D) mixed-mode SSM of the edge state in the plane Couette flow studied by \cite{kreilos_eckhardt_2012}. We show that this SSM-reduced model captures both the chaotic attractor and the laminar state of the flow with the SSM acting as an inertial manifold (\cite{foias_1988, liu_2023}). Our approach, therefore, also justifies the empirical low-dimensional model of \cite{kreilos_eckhardt_2012}.

\section{Setup}
We focus on the chaotic dynamics observed in the plane Couette flow, which is the incompressible flow in a channel between two infinite plates moving in opposite directions with velocity $\pm U$. The channel is defined as the domain $\Omega = \{(x,y,z) \in \mathbb{R}^3 \ : \ [0, L_x] \times [-h, h] \times [0, L_z] \}$, in which the velocity field $\mathbf{u} = [u, v, w](x,y,z,t)$ and the pressure $p$ satisfy the Navier--Stokes equations
\begin{equation}
\label{eq:ns}
    \frac{\partial \mathbf{u}}{\partial t} + \mathbf{u}\cdot \nabla \mathbf{u} = -\nabla p + \nu \Delta \mathbf{u}, \quad \nabla \cdot \mathbf{u} = 0,
\end{equation}
where $\nu$ is the kinematic viscosity. The main parameter of the problem is the Reynolds number defined as $\text{Re}=Uh/\nu$, where $2h$ is the distance between the moving walls. This also sets the relevant time unit as $h/U$. 

We work with a streamwise- and spanwise-periodic computational domain corresponding to the minimal flow unit studied by \cite{kreilos_eckhardt_2012} and \cite{kreilos_eckhardt_schneider_2014} and fix $h=1$, $L_x=2\pi$, and $L_z=\pi$. This domain is also comparable to those used by \cite{nagata_ThreedimensionalFiniteamplitudeSolutions1990}, \cite{page_KoopmanModeExpansions2019}, \cite{Linot_Graham_2023} and \cite{kaszas_DynamicsbasedMachineLearning2022}.

We simulate the flow using the open source \texttt{Channelflow} library (\cite{channelflow}), which employs a pseudo-spectral discretization with 32 Fourier modes in the streamwise and spanwise directions, and 33 Chebyshev modes in the wall-normal direction. The time evolution of the discretized PDE \eqref{eq:ns} can be interpreted as a trajectory of a finite-dimensional dynamical system,
\begin{equation}
\label{eq:dynsys}
    \dot{\mathbf{x}}=\mathbf{f}(\mathbf{x}), \ \mathbf{x}\in \mathbb{R}^N, \mathbf{f}\in \mathcal{C}^\infty,
\end{equation}
with $N\approx O(10^5)$. The laminar flow, which is a stable fixed point of \eqref{eq:dynsys}, is expressed as $u=y, v=w=0$. In the following, we focus on the dynamical system \eqref{eq:dynsys}, but note that the physical flow field $\mathbf{u}$ can always be recovered from $\mathbf{x}$ with the spatial discretization of \texttt{Channelflow}.

As shown by \cite{kreilos_eckhardt_2012} and \cite{kreilos_eckhardt_schneider_2014}, in this computational cell, the Nagata upper- and lower branch fixed points (\cite{nagata_ThreedimensionalFiniteamplitudeSolutions1990,gibsonEquilibriumTravellingwaveSolutions2009}) of \eqref{eq:dynsys} appear in a saddle-node bifurcation at $\text{Re}=163.8$. For higher Reynolds numbers, the upper branch fixed point (UB) undergoes a Hopf bifurcation, followed by a period doubling cascade, eventually leading to chaotic dynamics. 

The lower branch fixed point (LB) remains an edge state for a wide range of Reynolds numbers, and its codimension-one stable manifold forms the edge of chaos (\cite{wangLowerBranchCoherent2007, schneiderLaminarturbulentBoundaryPlane2008}). The chaotic attractor, which can be traced back to the UB, undergoes a boundary crisis (\cite{grebogi_1983}) and disappears at $\text{Re}=188.51$. After this point, the transient chaotic behavior associated with turbulence is generated by a chaotic saddle (\cite{lai_TransientChaos2011}) and hence has a finite lifetime (\cite{kreilos_eckhardt_schneider_2014}). 

We focus on a Reynolds number value before the boundary crisis, $\text{Re}= 187.8$, where a genuine chaotic attractor coexists with the stable laminar state. Furthermore, we also restrict the flow to the symmetry invariant subspace of the Nagata equilibria by solving \eqref{eq:ns} with the shift-reflect symmetries imposed. 

\subsection{Spectral submanifolds}
We aim to construct a simple reduced model that faithfully represents the bistability of \eqref{eq:dynsys}. To this end, we seek a low-dimensional invariant manifold containing the edge state, labeled $\mathbf{x}_{LB}$. The linearized dynamics around the edge state are governed by the Jacobian $D\mathbf{f}({\mathbf{x}_{LB}})$, whose eigenvalues are $\lambda_1,...,\lambda_N\in \mathbb{C}$.

Let us assume that the eigenvalues satisfy the nonresonance conditions
\begin{equation}
\label{eq:nonres}
    \sum_{j=1}^N m_j\lambda_j \neq \lambda_i \;\; m_j\in \mathbb{N},
\end{equation}
for $i=1,...,N$ and $\sum_{j=1}^Nm_j\geq2$. The conditions \eqref{eq:nonres} guarantee the applicability of the linearization theorem of \cite{sternberg_StructureLocalHomeomorphisms1958} for class $\mathcal{C}^\infty$ dynamical systems. This theorem guarantees the existence of a smooth transformation mapping \eqref{eq:dynsys} to its linearization around the edge state.

As shown by \cite{haller_NonlinearModelReduction2023}, this implies the existence of a family of SSMs tangent to any given spectral subspace of the linearized dynamics. Precisely one of these SSMs, the primary SSM, is $\mathcal{C}^\infty$ smooth, whereas all others have reduced differentiability. We target the slowest family of SSMs, tangent to the slowest spectral subspace spanned by eigenvectors of the linearization with eigenvalues closest to the imaginary axis. This is a normally attracting slow manifold if the remaining eigenvalues of $D\mathbf{f}_{\mathbf{x}_{LB}}$ have negative real parts.

\cite{liu_2023} and \cite{xu_2024} showed that slow mixed-mode SSMs tangent to both stable and unstable linear modes can contain the chaotic attractor of a dynamical system. Therefore, they often serve as inertial manifolds (\cite{foias_1988}). The existence of such mixed-mode SSMs of the Navier-Stokes equations can also be concluded from the results of \cite{buza_2024}.

\cite{kreilos_eckhardt_2012} constructed a 1D  Poincaré map by tracking the maxima of the kinetic energy signal on the chaotic attractor. Although such a construction only decreases the dimension of the system \eqref{eq:dynsys} by one, this 1D map already reveals periodic orbits in the attractor. These findings indicate that the attractor is likely low-dimensional and lies close to the edge of chaos. 
\section{Results}
\begin{figure}
    \centering
    \includegraphics[width=0.8\textwidth]{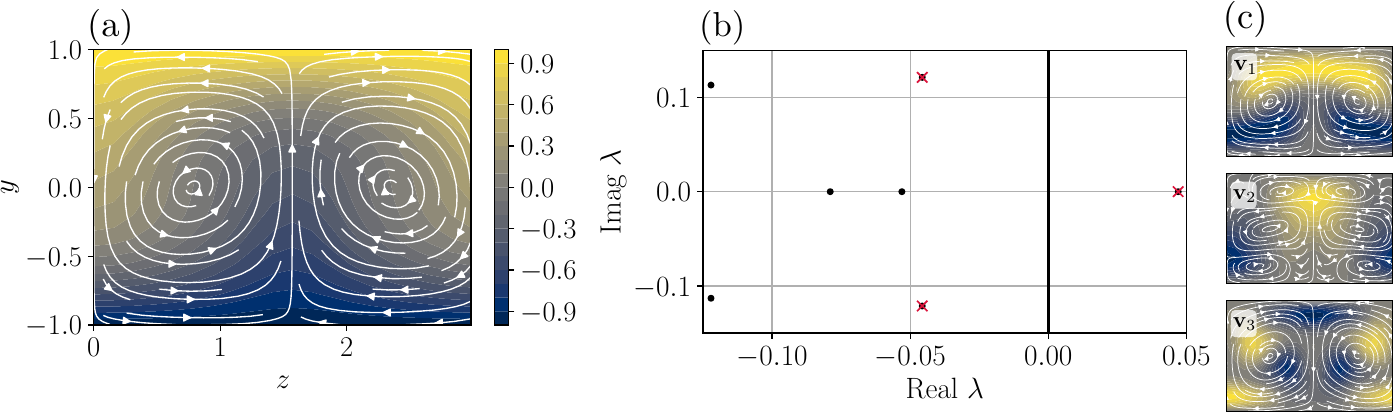}
    \caption{(a): Streamwise averaged velocity field of the edge state. The streamwise velocity is color coded, and the spanwise and wall-normal velocities are indicated as streamlines. (b): The spectrum of the edge state, computed by Channelflow. The eigenvalues associated with the spectral subspace $E$ spanned by $\mathbf{v}_{1,2,3}$, to which the SSM $\mathcal{W}(E)$ is tangent, are marked by red crosses. (c): Same as panel (a) for the eigenvectors $\mathbf{v}_{1,2,3}$.} 
    \label{fig:1}
\end{figure}

Figure \ref{fig:1} shows the edge state with its spectrum and its slowest eigenvectors labeled $\mathbf{v}_1, \mathbf{v}_2, \mathbf{v}_3$. These eigenvectors, corresponding to an unstable real eigenvalue and a pair of stable complex eigenvalues, span the slowest mixed-mode spectral subspace $E=\text{span}(\mathbf{v}_1, \mathbf{v}_2$, and  $\mathbf{v}_3)$. The slow SSM, $\mathcal{W}(E)$, is therefore to be constructed as a graph over these eigenvectors. 

We follow the \texttt{SSMLearn} algorithm of \cite{cenedese_DatadrivenModelingPrediction2022} to approximate the mixed-mode SSM, $\mathcal{W}(E)$, from data. Our training trajectories are initialized near the edge state and are first attracted to the slow SSM before converging to either the laminar state or the chaotic attractor. The DNS solver \texttt{Channelflow} also computes the eigenvectors $\mathbf{v}_{1,2,3}$ using the Arnoldi method, which we use to enforce the exact tangency between $E$ and $\mathcal{W}(E)$, i.e., we parameterize the SSM over the eigenmodes $\mathbf{v}_{1,2,3}$. 

The coordinate chart returning the reduced coordinates $\boldsymbol{\eta}$ is a projection to the eigenmodes \begin{equation}\label{eq:chart}
\boldsymbol{\eta}=\mathbf{V}^T\mathbf{x},   
\end{equation} where $\mathbf{V}$ is the matrix containing $\mathbf{v}_{1,2,3}$. 
In addition, the eigenvectors allow us to ensure that the training trajectories lie close to the SSM. Specifically, we initialize the trajectories on the spectral subspace $E$, which are velocity fields of the form
\begin{equation}\label{eq:ics}
\mathbf{x}=\mathbf{x}_{LB}+\alpha_1\mathbf{v}_1 + \alpha_2\mathbf{v}_2  + \alpha_3\mathbf{v}_3,    
\end{equation}
with small coefficients $\alpha_{1,2,3}$. The sign of $\alpha_1$ decides on which side of the edge of chaos the initial condition lies. Our approach, therefore, is not strictly data-driven, but data-assisted, using the terminology of \cite{Cenedese2025}.

\begin{figure}
    \centering
    \includegraphics[width=0.75\textwidth]{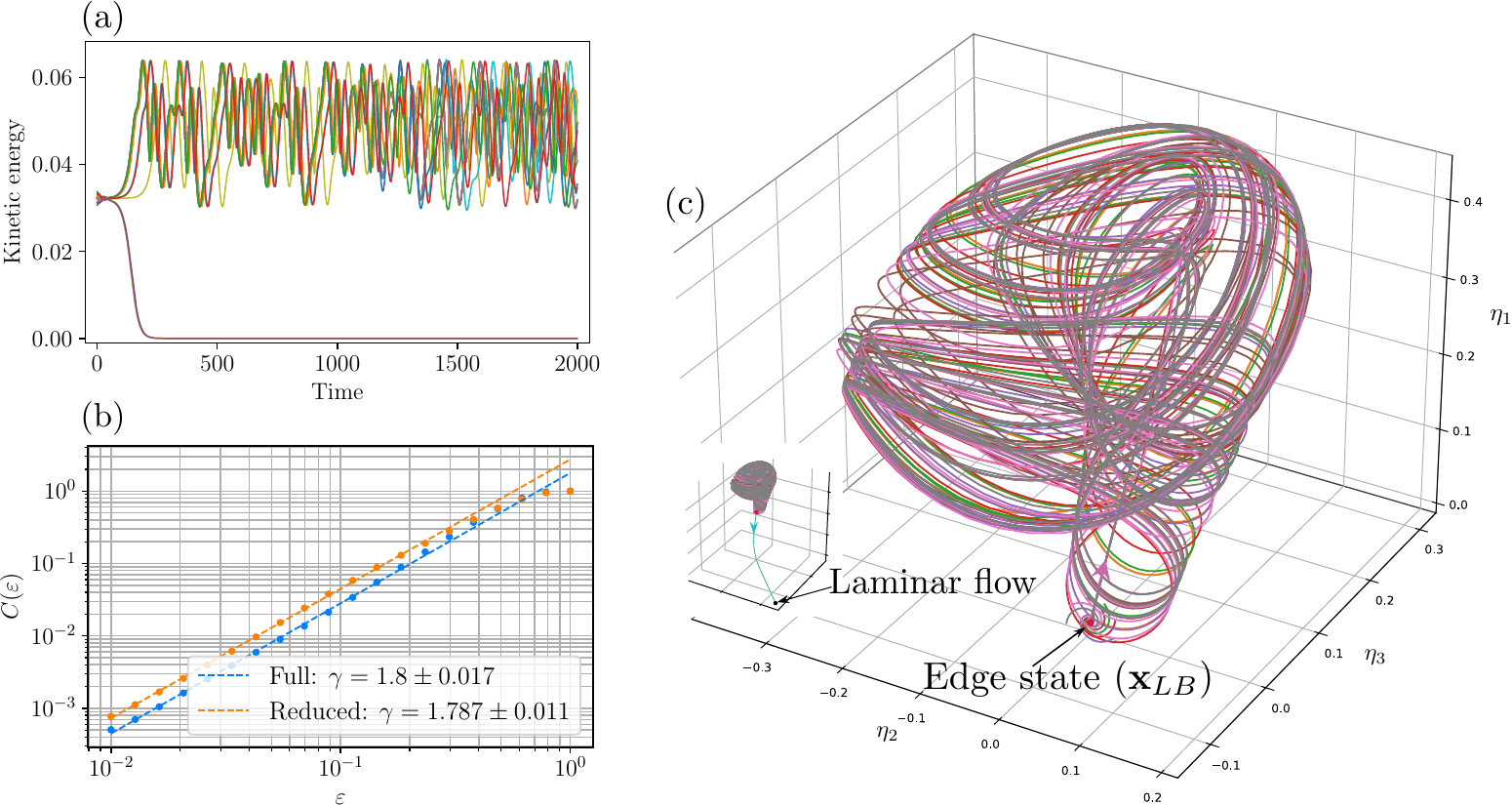}
    \caption{(a): The average kinetic energy along training trajectories. (b): Correlation dimension estimation based on \eqref{eq:corr_dim}, in the full phase and the reduced phase space. The corresponding power-law fits are shown in blue and orange, respectively. (c): The reduced coordinates of the same trajectories as they converge to the chaotic attractor. The inset shows the laminarizing trajectories as well.}  
    \label{fig:2}
\end{figure}

We advect 12 initial conditions \eqref{eq:ics} using \texttt{Channelflow} up to time $T_{max}=2,000$. In total, this results in $N_d=24,000$ data points in the training set. The kinetic energy $E=\frac{1}{2}\|\mathbf{u}\|^2$ of these trajectories, averaged over the domain $\Omega$, is shown in Fig. \ref{fig:2}a. The reduced coordinates of the training trajectories, given by \eqref{eq:chart}, are shown in Fig. \ref{fig:2}c. On the chaotic side of the edge, we see saddle-spiral type dynamics, reminiscent of Shilnikov-type chaos (\cite{Shilnikov1998}), while on the other side, the dynamics are essentially 1D, leading to rapid laminarization. 

To find further evidence of the low-dimensionality of the chaotic attractor, we estimate its correlation dimension using the training trajectories. As defined by \cite{grassberger_1983}, the correlation dimension is the scaling exponent $\gamma$ in the relation
\begin{equation}
\label{eq:corr_dim}
    C(\varepsilon) \sim \varepsilon^{\gamma}, \text{ as } \varepsilon\to 0,
\end{equation}
where $C(\varepsilon)$ is the correlation sum, i.e., the number of pairs of points in the attractor that are separated by a distance less than $\varepsilon>0$. 

We computed the correlation sum $C(\varepsilon)$ in the full, $N$-dimensional phase space and in the reduced, 3D phase space. The results are shown in Fig. \ref{fig:2}b. A linear fit on a logarithmic scale returns a correlation dimension of approximately $\gamma=1.8$ in both cases. Therefore, the chaotic attractor is indeed low-dimensional and no topological information is lost by restricting the dynamics to the SSM.

\begin{figure}
    \centering
    \includegraphics[width=0.75\textwidth]{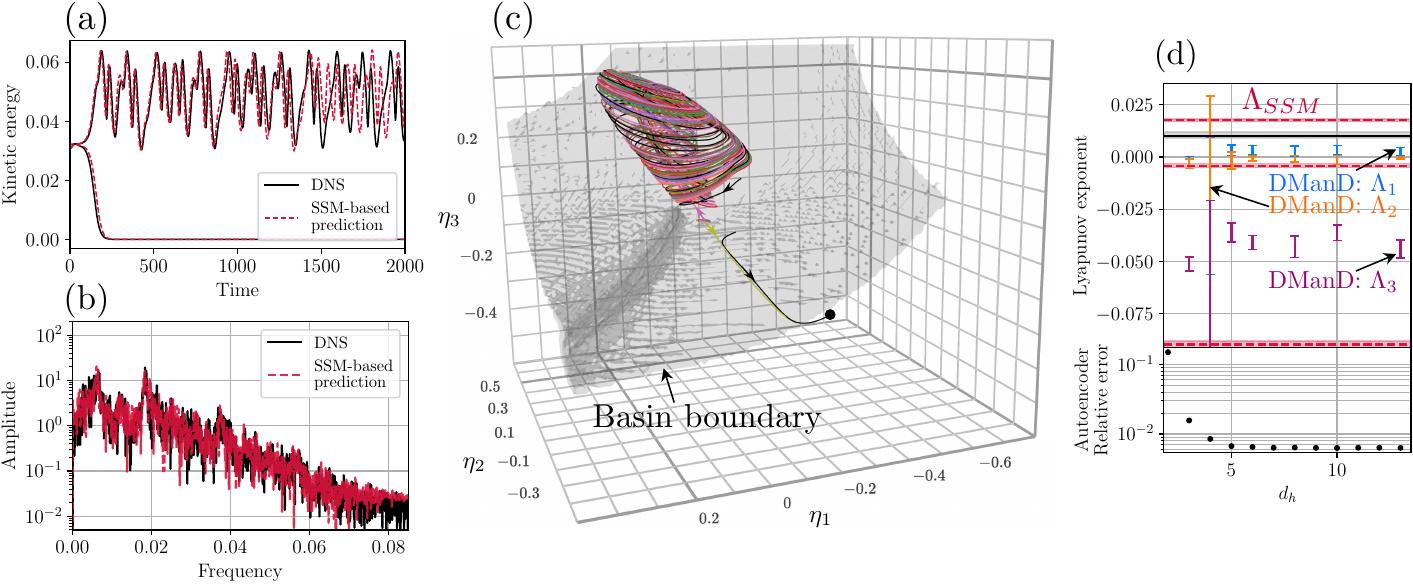}
    \caption{(a): Model predictions of test trajectories. (b): Power spectral densities computed from a chaotic kinetic energy signal in the full model and the SSM-reduced model. (c): A subset of the edge of chaos, constructed as the boundary of the basins of attraction in the reduced model, is shown with the training trajectories. Trajectories initialized on either side of the edge of chaos are also indicated as black lines.  (d): Three leading Lyapunov exponents and relative reconstruction errors of the \texttt{DManD} model for various dimensions of the latent space $d_h$. The gray line indicates the leading Lyapunov exponent of the true model. Red lines show the Lyapunov exponents of the SSM-model.}
    \label{fig:3}
\end{figure}

Having defined the coordinate chart \eqref{eq:chart}, we proceed by defining the parametrization of the SSM. We first shift the coordinates to place the edge state $\mathbf{x}_{LB}$ at the origin. We then seek the parametrization in the following polynomial form 
\begin{equation}\label{eq:parametrization}
    \mathbf{x}=\mathbf{V}\boldsymbol{\eta} + \sum_{n=2}^{M_p}\sum_{i+j+k=n}\mathbf{W}_{ijk}\eta_1^i \eta_2^j \eta_3^k,
\end{equation}
where $\mathbf{W}_{ijk}$ are the coefficients of the monomial terms. We use \texttt{SSMLearn} to determine the polynomial coefficients by minimizing the error between the training data and \eqref{eq:parametrization}. Using cross validation, we determine that $M_p=5$ results in the optimal reconstruction error. For more information, we refer to the JFM Notebook accompanying Fig. \ref{fig:3}. 

After identifying the geometry of the SSM, $\mathcal{W}(E)$, we approximate the reduced dynamics. We model it as a discrete dynamical system given by the 3D iterated map
\begin{equation}
    \label{eq:reduced_dyn_def}
    \boldsymbol{\eta}_{n+1}=\mathbf{F}(\boldsymbol{\eta}_n), \quad \mathbf{F}:\mathbb{R}^3\to \mathbb{R}^3,
\end{equation}
where $\boldsymbol{\eta}_n$ is the reduced state at time step $n$.

Although the flow of \eqref{eq:dynsys} is fundamentally continuous in time, we use a sampling time of $\Delta t=1$ to obtain the reduced map \eqref{eq:reduced_dyn_def}. Since the sampling time is small compared to other relevant time scales, we treat the trajectories of \eqref{eq:reduced_dyn_def} as if they were continuous in time by linearly interpolating between discrete time steps whenever necessary. 

A polynomial approximation of the reduced dynamics is insufficient for capturing the chaotic dynamics, as reported by \cite{liu_2023} and \cite{xu_2024}.  Instead, we use the \texttt{SSMLearn} algorithm with an alternative interpolation method. In particular, we approximate the dynamics as a linear combination of radial basis functions (RBFs) (\cite{Buhmann_2003}).

The map \eqref{eq:reduced_dyn_def} is then approximated in the form
\begin{equation}
    \label{eq:reduced_dyn}
    \mathbf{F}(\boldsymbol{\eta})=\sum_{i=1}^{N_d}\mathbf{c}_i k(\|\boldsymbol{\eta}_i-\boldsymbol{\eta}\|),
\end{equation}
where $\boldsymbol{\eta}_i$, $i=1,...,N_d$ are the training data points, and $k:\mathbb{R}\to\mathbb{R}$ is a radial function. We use the multiquadric kernel $k(r)=\sqrt{r^2 + \epsilon^2}$ with $\epsilon=10^{-7}$.  
The coefficients $\mathbf{c}_i\in \mathbb{R}^3$ are determined by linear regression, minimizing the squared error between the training data and the RBF approximation. 

The reduced dynamics \eqref{eq:reduced_dyn} can be used to predict the time evolution of previously unseen initial conditions, i.e., test trajectories. We show two such predictions in Fig. \ref{fig:3}a. The initial conditions are close to the edge state, but on opposite sides of the edge of chaos. The laminarizing trajectory is accurately predicted over the whole time interval. As expected, predictions of the chaotic trajectory are accurate only on shorter timescales. This is due to the sensitivity of chaotic trajectories to initial conditions. 

We can also use the reduced dynamics \eqref{eq:reduced_dyn} to compute the Lyapunov exponents of the chaotic attractor. The Lyapunov exponents are defined as the average exponential growth rate of perturbations along the chaotic attractor. We compute the Lyapunov exponents by iterating the equation of variations. Computing the Jacobian of \eqref{eq:reduced_dyn_def} directly, we find that the linearized flow map $\boldsymbol{\Phi}_n=\frac{\partial \boldsymbol{\eta}_n}{\partial \boldsymbol{\eta}_0}$ with $\boldsymbol{\eta}_0$ denoting the initial condition, is governed by

\begin{equation}
        \boldsymbol{\Phi}_{n+1} = \mathbf{DF}(\boldsymbol{\eta}_n)\boldsymbol{\Phi}_n, \quad
    \mathbf{DF}(\boldsymbol{\eta})=\sum_{i=1}^{N_d}\mathbf{c}_i\otimes \frac{\partial k}{\partial \boldsymbol{\eta}}(\|\boldsymbol{\eta}_i-\boldsymbol{\eta}\|) = \sum_{i=1}^{N_d}\frac{1}{\sqrt{\|\boldsymbol{\eta}_i-\boldsymbol{\eta}\|^2+\epsilon^2}}\mathbf{c}_i\otimes (\boldsymbol{\eta}_i-\boldsymbol{\eta}),
\end{equation}
where $\otimes$ denotes the tensor product. If they exist, the $n\to \infty $ limits of the singular values of $\boldsymbol{\Phi}_n$ define the Lyapunov exponents of the attractor. 
We find that the largest Lyapunov exponent is $\Lambda_{SSM}= 0.017\pm 0.002$, which is larger than what we estimated from the rate of divergence of initially close-by trajectories of the full system, which is $\Lambda_{DNS}=0.009 \pm 0.002$. However, we note that the uncertainty of $\Lambda_{DNS}$ is likely underestimated because we do not have access to the Jacobian of the full system.

We can also compute relevant statistics on the chaotic attractor. In Fig. \ref{fig:3}b we show the broad power spectral density obtained from the Fourier transform of the kinetic energy signal, characteristic of chaotic dynamics. A comparable time series of the full model yields a similar spectrum, with inaccuracies visible only towards high frequencies. 

We further investigate the chaotic dynamics on the SSM by constructing the basins of the two coexisting attractors. We fill the domain of the reduced phase space shown in Fig. \ref{fig:3}c with a total of $10^6$ initial conditions and iterate them forward under the reduced dynamics \eqref{eq:reduced_dyn} for $T = 1,000$ steps. We then record the initial conditions of the laminarizing trajectories. This yields a characteristic function of the basin of attraction of the laminar state, with the basin of the chaotic attractor obtained as its complement. 

Instead of visualizing the characteristic function directly in the reduced phase space, we construct the boundary between the two basins. We define this basin boundary as the level set of the characteristic function at the value of $0.5$. The largest connected component of this level set, approximating the stable manifold of the edge state, is shown in Fig. \ref{fig:3}c. 
This set is a 2D intersection of the edge of chaos with the SSM $\mathcal{W}(E)$. \cite{Kaszas_Haller_2024} recently obtained a 1D SSM-edge intersection for turbulent pipe flow.

We have also employed the recent neural ODE-based data-driven modeling method of \cite{Linot_Graham_2023} using our training data. The \texttt{DManD} algorithm reduces the dynamics to an inertial manifold by an initial linear projection to 500 POD modes, followed by the application of an autoencoder. This maps the data to a latent space of dimension $d_h$, parametrizing the inertial manifold. The reduced dynamics are then modeled as a neural ODE trained on the dynamics of the latent variables. We used the code published by \cite{Linot_Graham_2023} to train \texttt{DManD} models with $d_h = 3, ..., 13$, repeating the training 10 times for each latent dimension.

The autoencoder reconstruction error saturates already for $d_h>5$, but the models representing the reduced dynamics often produce non-chaotic dynamics for lower dimensions. We see consistently chaotic models for $d_h\geq10$, but their leading Lyapunov exponent is slightly lower than expected. Since \texttt{DManD} was proposed to model purely chaotic behavior, the discrepancy is likely due to the simultaneous presence of chaotic and laminar trajectories in our training data. 

We summarize our results in Fig. \ref{fig:3}d, with the details of the training and evaluation of the \texttt{DManD} models reported in the accompanying JFM Notebook. We found that accurately modeling systems with multiple coexisting attractors is currently out of reach for \texttt{DManD}, and hence likely for other neural network-based modeling approaches as well. Resolving these issues will require future research and adjustments to the network architecture. 
\subsection{Poincaré map}
Once the reduced dynamics \eqref{eq:reduced_dyn} are obtained, we can reduce the model dimension even further by defining a Poincaré map (\cite{guckenheimerNonlinearOscillationsDynamical1983}). Since Poincaré maps are defined for continuous-time dynamical systems, we need to interpret \eqref{eq:reduced_dyn} as the $\Delta t=1$-sampled equivalent of a continuous-time system generated by a vector field. For instance, if we assume \eqref{eq:reduced_dyn} was obtained by Euler's method with a time step of $\Delta t=1$, i.e., the governing ODE would be $\dot{\boldsymbol{\eta}} = \mathbf{F}(\boldsymbol{\eta})-\boldsymbol{\eta}$. 

\begin{figure}
    \centering
    \includegraphics[width=0.75\textwidth]{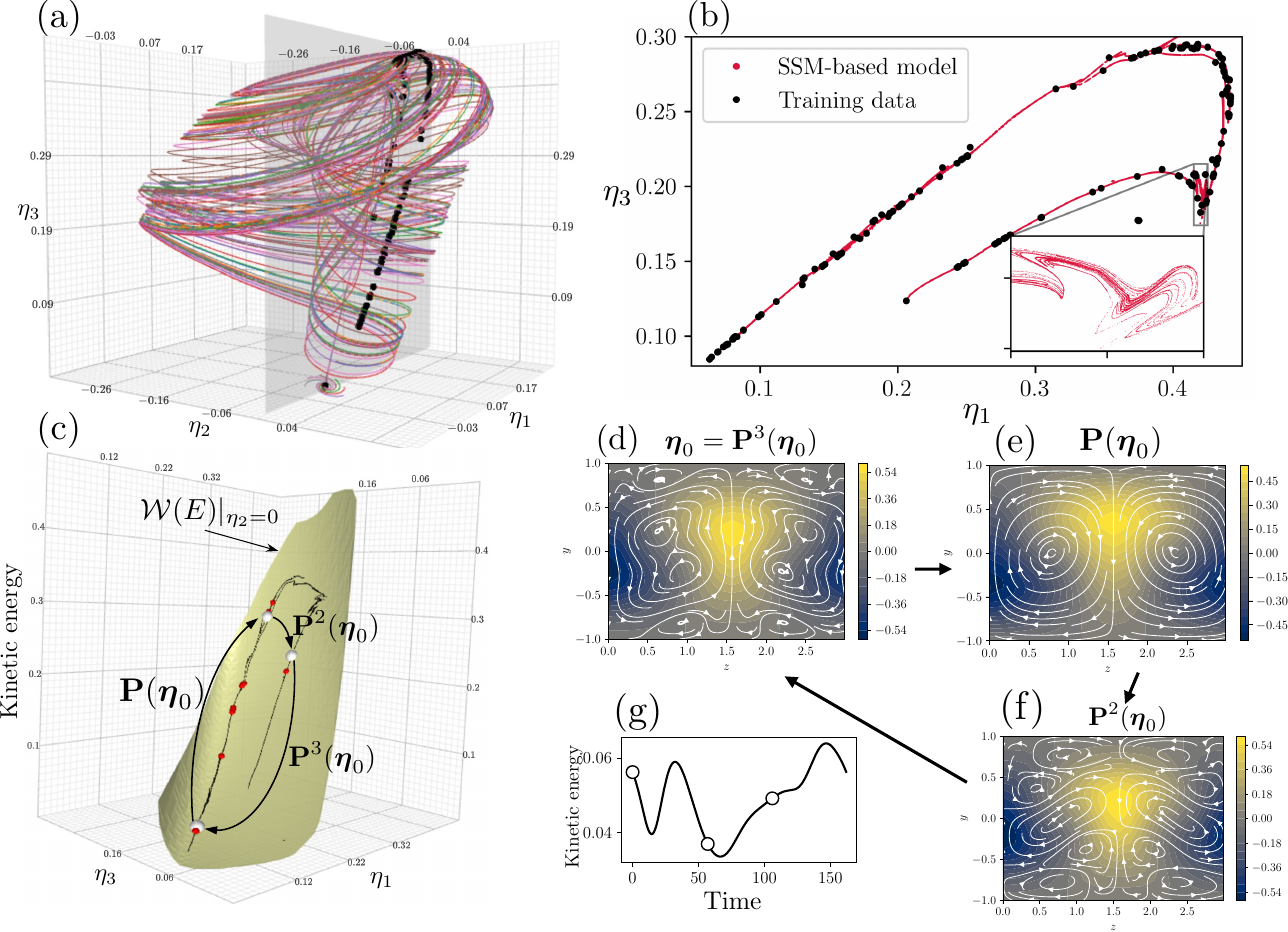}
    \caption{(a): Black dots indicate the Poincaré section of the chaotic attractor with the $\eta_2=0$ plane (gray). (b):  Black dots indicate intersections computed based on the training trajectories. Red dots are iterations of the SSM-reduced Poincaré map. (c): Poincaré section of the SSM $\mathcal{W}(E)$, containing the chaotic attractor (black) and periodic orbits (red). A period-3 orbit is highlighted in white, with the flow fields shown in panels (d-f) using the same visualization as Fig. \ref{fig:1}. The kinetic energy of this orbit is shown in panel (g).}  
    \label{fig:4}
\end{figure}

However, instead of solving the ODE, we formally define the continuous-time flow of the reduced dynamics by linearly interpolating between successive time steps. The Poincaré (or first-return) map $\mathbf{P}(\boldsymbol{\eta})$ is then defined by successive intersections of the trajectory with a plane in the reduced phase space. We select the plane to be $\eta_2=0$ and record intersections with $({\eta}_2)_{n+1}>0>({\eta}_2)_{n}$. Figure \ref{fig:4}a shows the intersection of the chaotic attractor with $\eta_2=0$. 

Using the reduced dynamics, the map $\mathbf{P}$ can be evaluated arbitrarily many times to obtain a high-resolution representation of the chaotic attractor. This is shown in Fig. \ref{fig:4}b, which features a total of $10^6$ intersection points generated by $\mathbf{P}$, revealing a much richer structure than the limited number of sample points obtained from the training trajectories. Particularly, the SSM-reduced model revealed the small-scale fractal structure of the attractor in the inset of Fig. \ref{fig:4}b. Visually, the attractor resembles the attractor of the Hénon-map, as remarked also by \cite{kreilos_eckhardt_2012}.

The Poincaré section of the SSM, $\mathcal{W}(E)$, denoted $\mathcal{W}(E)|_{\eta_2=0}$, can be computed by simply setting $\eta_2=0$ in the parametrization \eqref{eq:parametrization}. In Figure \ref{fig:4}c, we visualize the attractor of the Poincaré map on the SSM, by computing the kinetic energy of the flow fields.

We can also use the low-dimensional Poincaré map on the SSM to extract previously unobserved features of the chaotic dynamics. In particular, since the chaotic attractor is the closure of infinitely many unstable periodic orbits (\cite{guckenheimerNonlinearOscillationsDynamical1983}), we can use the Poincaré map to find some of these underlying orbits. A similar approach was used by \cite{kreilos_eckhardt_2012} and \cite{Linot_Graham_2023} and is in contrast to the usual methodology of finding periodic orbits to approximate the chaotic dynamics (\cite{page_norgaard_2024}). 

\cite{Linot_Graham_2023} search for periodic orbits of the reduced dynamics directly, i.e. they identify initial conditions $\boldsymbol{\eta}_0$ for which $\boldsymbol{\eta}(T;\boldsymbol{\eta}_0)=\boldsymbol{\eta}_0$ holds. Note, however, that this requires finding an appropriate period $T\in\mathbb{R}^+$ simultaneously. 

The Poincaré map $\mathbf{P}$ bypasses the problem of finding the period, because periodic orbits appear as fixed points of iterates of $\mathbf{P}$. We therefore search for solutions of the equation $\mathbf{P}^n(\boldsymbol{\eta})-\boldsymbol{\eta} = \boldsymbol{0}$, while systematically increasing $n$. We use a Newton-Raphson-like method to find these roots.

We initialize the root finding method with $n=1,...,5$ from random initial conditions distributed along the attractor. Upon successful identification of a periodic orbit, the initial condition is then provided as an initial guess to \texttt{Channelflow}'s Newton-Krylov solver (\cite{viswanath_RecurrentMotionsPlane2007}). Although we did not include any periodic orbits in the training data, we successfully found eight periodic ECSs of the full system using the initial guesses from the reduced model. One of these orbits is shown in Fig. \ref{fig:4}d-g. 

We note that some of these ECSs have already been found by \cite{kreilos_eckhardt_2012}, but we were able to find others as well. We emphasize that our goal was not to compile an extensive library of periodic orbits of this particular system, but to demonstrate that an SSM-based reduced model already captures the essential chaotic dynamics. Indeed, the reduced dynamics enables the efficient sampling of the natural measure of the chaotic attractor to compute the probability distribution of quantities of interest, such as the spectrum of the kinetic energy in Fig. \ref{fig:3}b.


\section{Conclusions and discussion}
We have applied the theory of spectral submanifolds to plane Couette flow in the parameter regime supporting a chaotic attractor. Although the flow is simulated in a minimal flow unit at a relatively low Reynolds number, successful modeling of such simple, but physically relevant systems is an important step towards modeling turbulence.

We restricted the flow to the slowest 3D mixed-mode SSM of the edge state, which functions as an inertial manifold. Our very low-dimensional reduced model is given by an iterated map interpolated using radial basis functions. Although contemporary machine learning methods, such as neural ODEs (\cite{linot_DeepLearningDiscover2020}), could also be implemented to model the dynamics, we found that simpler interpolation methods already deliver excellent accuracy.

Compared to these machine learning methods, the classical function approximation used here has the advantage that it does not require any iterative training. This makes the models less data-hungry, faster to compute, and their performance more predictable. This is due to the exact mathematical background supporting the existence and smoothness properties of the underlying SSMs. In addition, we found that the state-of-the-art neural network-based model, \texttt{DManD} (\cite{Linot_Graham_2023}), may have difficulties modeling the dynamics due to the coexisting chaotic and laminar attractors. In order for the neural ODE model to produce chaotic dynamics consistently, larger-dimensional inertial manifolds were required. 

We have also defined a 2D Poincaré map associated with the SSM-reduced model to extract previously unseen characteristics of the dynamics. In particular, the Poincaré map revealed the fractal structure of the chaotic attractor and facilitated the search for periodic orbits. Although \cite{kreilos_eckhardt_2012} inferred a similar 1D map based on simulated time histories of the kinetic energy to find periodic orbits, our approach makes a rigorous connection between the dynamics in the full phase space and a 2D Poincaré section in the reduced phase space. 

Our SSM-reduced model has proven to generalize effectively and represent the chaotic attractor accurately. Therefore, we believe that constructing SSM models, attached to known ECS, is a promising method for modeling other flows, too. In particular, they allow efficient sampling of the probability distributions of physical variables without relying on periodic orbit theory (\cite{ChaosBook}).
\backsection[Acknowledgements]{We are grateful to Alec Linot and Michael Graham for making their DManD code available and for helpful discussions and useful advice on its implementation.} 
\backsection[Declaration of Interests]{The authors report no conflict of interest.}

\bibliographystyle{jfm}

\end{document}